\documentclass[prl,aps, superscriptaddress,twocolumn]{revtex4-2}
\usepackage{amsmath,amssymb}
\usepackage{graphicx}
\usepackage{wasysym}
\usepackage{amsfonts}
\usepackage{bm}
\usepackage{enumerate}
\usepackage[dvipsnames]{xcolor}
\usepackage[resetlabels]{multibib}
\usepackage{epstopdf}
\usepackage{latexsym}
\usepackage{multirow}
\usepackage{booktabs}
\usepackage[breaklinks,colorlinks = true,linkcolor = red,urlcolor=cyan,citecolor=red]{hyperref}
\usepackage[caption=false,singlelinecheck=false]{subfig}

\usepackage{times}
\newcommand{\bea}{\begin{eqnarray}}
\newcommand{\eea}{\end{eqnarray}}
\newcommand{\be}{\begin{eqnarray}}
\newcommand{\ee}{\end{eqnarray}}
\newcommand{\bw}{\begin{widetext}}
\newcommand{\ew}{\end{widetext}}
\newcommand{\nn}{\nonumber}

\newcommand{\la}{\langle}
\newcommand{\ra}{\rangle}

\newcommand{\tbf}{\textbf}

\makeatletter
\newcommand*{\sumcirclearrowleft}{%
  \DOTSB
  \mathop{
    \mathchoice
      {\rlap{\kern.25em\rotatebox[origin=c]{-90}{$\circlearrowleft$}}{\sum}}
      {\vcenter{\rlap{\kern.2em\rotatebox[origin=c]{-90}{$\scriptscriptstyle\circlearrowleft$}}}{\sum}}
      {\sum}{\sum}
  }\slimits@
}

\newcommand*{\sumcirclearrowright}{%
  \DOTSB
  \mathop{
    \mathchoice
      {\rlap{\kern.25em\rotatebox[origin=c]{90}{$\circlearrowright$}}{\sum}}
      {\vcenter{\rlap{\kern.2em\rotatebox[origin=c]{90}{$\scriptscriptstyle\circlearrowright$}}}{\sum}}
      {\sum}{\sum}
  }\slimits@
}
\makeatother

\usepackage{soul}

\makeatletter
\def\SOUL@ulstunderline#1{{%
    \setbox\z@\hbox{#1}%
    \dimen@=\wd\z@
    \dimen@i=\SOUL@uloverlap
    \advance\dimen@2\dimen@i
    \rlap{
        \null
        \kern-\dimen@i
        \SOUL@ulcolor{\SOUL@ulleaders\hskip\dimen@}%
    }%
    \SOUL@stpreamble
    \rlap{%
        \null
        \kern-\dimen@i
        \SOUL@ulcolor{\SOUL@ulleaders\hskip\dimen@}%
    }%
    \unhcopy\z@
}}
\def\SOUL@ulsteverysyllable{%
    \SOUL@ulstunderline{%
        \the\SOUL@syllable
        \SOUL@setkern\SOUL@charkern
    }%
}
\def\SOUL@ulstsetup{%
  \SOUL@ulsetup
  \let\SOUL@everysyllable\SOUL@ulsteverysyllable
}
\DeclareRobustCommand*\textulst{\SOUL@ulstsetup\SOUL@}
\makeatletter

\begin{document}
\title{Fractionalization induced structural domain patterns in U(1) quantum spin liquids }

\author{Hyeok-Jun Yang}
\email{yang267814@kaist.ac.kr}
\affiliation{Department of Physics, Korea Advanced Institute of Science and Technology, Daejeon, 34141, Korea}

\author{Eun-Gook Moon}
\email{egmoon@kaist.ac.kr}
\affiliation{Department of Physics, Korea Advanced Institute of Science and Technology, Daejeon, 34141, Korea}

\author{SungBin Lee}
\email{sungbin@kaist.ac.kr}
\affiliation{Department of Physics, Korea Advanced Institute of Science and Technology, Daejeon, 34141, Korea}
\date{\today}

\begin{abstract}
The emergence of fractionalized quasiparticles in quantum spin liquids has served a wealth of unconventional phenomena in frustrated magnets.
In our work, we explore the various domain patterns of such fractionalized quasiparticles, especially focusing on charge defects in U(1) quantum spin liquids.
We claim that emergent long range interaction between charge defects leads to characteristic structures with distinct length scales, where they can be controlled via the ratio of interaction strengths. 
In this context, the spin ice phase is the dilute gas of weakly interacting charges, whereas, the macroscopic population of charge defects naturally develops charge ordering for large Coulomb interaction limit. 
Interestingly, we find that the competing spin interactions could naturally give rise to stabilize the mosaic structure of charge defects in the absence of uniform ordering. They are characterized by liquid-like correlations having a finite length scale. 
The emergence of such intermediate order in the mosaic structure is confirmed by both dynamical and static correlations.
By establishing the microscopic spin Hamiltonian, we also present the distinctive signatures in static spin correlation to detect such spatial structure of charge defects. 
We speculate that the domain pattern of defect population might be a potential hallmark to reveal unusual dynamical properties observed in spin liquids.
\end{abstract}
\maketitle
\label{sec:Introduction}
{\textbf{\textit {  Introduction ---}}}
One of the intriguing phenomena of frustrated interactions is the emergence of amorphous orders in a broad range of physical systems \cite{PhysRevB.26.325, doi:10.1126/science.267.5197.476, RevModPhys.83.587}.
The frustration typically prevents the uniform extension of locally preferred order, instead it might develop complex patterns of rich domains when the characteristic length takes on mesoscopic scale. 
A prominent example is the strip structure of charge arrays in doped Mott insulators in which the microphase separation occurs into hole-rich and hole-poor regions \cite{EMERY1993597, Tranquada1995, Kivelson1998}.
While the domain pattern is extensively studied in glass or liquid-like phases such as super-cooled liquids \cite{doi:10.1063/1.468414, KIVELSON199527, 10.1143/PTP.126.289, Tarjus_2005}, liquid crystals \cite{Carlson1988} and micellar solutions \cite{leibler1980theory,Matsen1996}, the comprehensive understanding of these complex liquids is a far-reaching goal. 
Specifically, the emergence of intermediate length scale is rarely expected in magnetic systems with short-range interactions.

Recently, it has been pointed out that the fractionalization in quantum spin liquid might be a key ingredient to generate the structural frustration \cite{PhysRevLett.104.107201, doi:10.1073/pnas.1317631111, Rau2016, PhysRevB.94.104416, Hart2021, PhysRevResearch.4.033159}.
Among them, a U(1) quantum spin liquid has gained attention as a unique platform for searching unconventional phenomena, such as inequilibrium dynamics and structural glassiness.
Particularly in spin ice, the fractionalized quasiparticle emerges as a point-like defect violating the divergenceless condition, so called a monopole defect \cite{PhysRevLett.95.217201, Castelnovo2008,Jaubert2009,Morris2009,Fennell2009}. Such monopole charge cannot be created or annihilated alone and the population is managed by the analogue of open Dirac strings.
It manifests the emergent Coulomb interaction and imposes a long-range frustration between the magnetic monopoles.
Thus, the mobile defects are kinetically constrained, which leads to intriguing dynamical phenomena such as the slow relaxation. 
Moreover, one may expect that the long-range Coulomb interaction leads to unconventional orderings of charge defect which has not been explored in conventional magnets.

In this paper, we study domain patterns of charge defect clustering on a 3-dimensional U(1) spin liquid. 
On top of the inborn long-range Coulomb interaction, the short-range attraction between same-charge defects is considered.  While the short-range attraction is built from spin exchanges \cite{Rau2016, PhysRevB.94.104416, PhysRevB.98.144446}, the strength of long-range frustration can be controlled by magnetic disorder and dipolar interactions \cite{Castelnovo2008,PhysRevLett.110.107202}.
Then, the ground state generically stabilizes the formation of locally correlated clusters of same-charge defects \cite{PhysRevLett.83.472, PhysRevLett.85.836,PhysRevE.62.7781, PhysRevE.65.065103,PhysRevLett.72.1918}. 
At the same time, the short-range interaction competes with the long-range frustration which prevents the uniform extension of local domains.
The competing tendency acting on different length scales leads to the emergence of intermediate length and spatial inhomogeneity.

We estimate the ground state phase diagram of domain patterns of fractionalized charge defect,  
by exemplifying a three dimensional cubic lattice case.
 For large Coulomb term, the typical size of clusters is relatively small with a characteristic momentum $\tbf{R}=\pi(1,1,1)$, similar to the Néel order.
As the short-range attraction increases, the cluster size grows and the modulation considerably deviates from $\tbf{R}$.
By manipulating cluster sizes, we show that various patterns of translation symmetry breaking are realized beyond the uniform and Néel-type orders.
We show that the charge correlation in the domain patterns is closely related to the spin correlation in elastic scattering experiments.
Remarkably, the packing of clusters for large attraction develop the mosaic structure with an extremely slow relaxation without long-range order. 
The half-moon shape in static spin structure signifies that the spatial structure is inhomogeneous and is an analogue of isotropic liquids correlated over the intermediate length scale.
We expect that spin liquids supported by quantum tunneling between cluster phases would manifest unique features of spatial inhomogeneity and liquid-like correlation.
Our study gives general insights into unusual static and dynamical properties in spin liquids.

\label{sec:Effective model}
{\textbf{\textit {Effective model ---}}}
We consider the low-energy effective action, $\mathcal{S}[Q_{\tbf{r}}]$ for the charge defect \cite{EMERY1993597,PhysRevLett.72.1918}, including both the short-range and long-range interactions in a U(1) quantum spin liquid.
\bea
\frac{1}{\beta}\mathcal{S}[Q_\tbf{r}]
=
\frac{\mathcal{R}}{2}\sum_{\tbf{r}}Q_\tbf{r}^2
-\mathcal{K}\sum_{\la \tbf{r}\tbf{r}'\ra}Q_{\tbf{r}}Q_{\tbf{r}'}
+\frac{\mathcal{U}}{2}\sum_{\tbf{r}\neq\tbf{r}'}\frac{Q_{\tbf{r}}Q_{\tbf{r}'}}{|\tbf{r}-\tbf{r}'|}, \;\;
\label{eq:Q_eff}
\eea
where $\mathcal{R}, \mathcal{K},\mathcal{U}>0$ and $\beta$ is the inverse temperature.
The defect charge is discrete, $Q_\tbf{r}=0, \pm 1, ...$ and placed on a 3-dimensional simple cubic lattice site $\tbf{r}$. Here, $|\tbf{r}-\tbf{r}'|$ is the Euclidean distance between $\tbf{r}$ and $\tbf{r}'$ in unit of the lattice spacing. 
Later, the microscopic spin Hamiltonian will be formulated, whose effective action of defects results in Eq. (\ref{eq:Q_eff}).
We emphasize that the Coulomb interaction between defects is not screened by anisotropic nature and it decays much slower than the dipole-dipole interaction. 
In addition, the non-local characteristics of defects result in distinct phases beyond the traditional magnets with spontaneous symmetry breaking.

In Eq, (\ref{eq:Q_eff}), the first term imposes an energy cost for creating charge defect at site $\tbf{r}$, which controls the defect concentration.
The second term represents a short-range attraction between same-charge defects.
The locally correlated cluster consists of positive (or negative) charges and grows as $\mathcal{K}$ increases.
Meanwhile, the long-range Coulomb interaction with the strength, $\mathcal{U}$, forbids the the uniform order with a momentum $\tbf{k}=0$, otherwise the third term diverges as $\sim 1/|\tbf{k}|^2$.
As a result, the competing interactions in Eq. (\ref{eq:Q_eff}) generically stabilizes a inhomogeneous mixture of rich domains of empty, positive- and negative charges.

\begin{figure}[t!]
{\label{fig:Q_pair}
\includegraphics[width=0.27\textwidth]{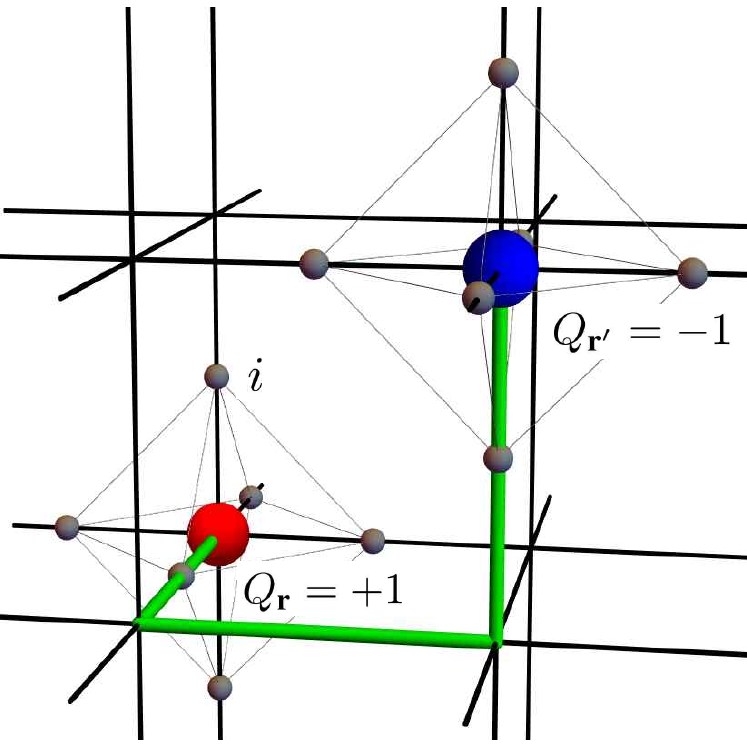}}\quad
\caption{
The pair creation (or annihilation) process of positive (red sphere) and negative (blue sphere) charges responsible for the spatial structure and dynamics of Eq. (\ref{eq:Q_eff}). 
Each process is implemented by the Dirac string (green cylinder) of longitudinal modes, $S_{i}^{(L)}$ to keep the global neutrality, Eq. (\ref{eq:Q_total}). The spin degree of freedom is defined on the midpoint $i$ (gray sphere) of the link, or equivalently the vertex of the octahedron surrounding the cubic site. 
}
\label{fig:Q_pair}
\end{figure}

On top of the competing interactions, we note that the long-range Coulomb interaction naturally takes place when the degrees of freedom, $Q_\tbf{r}$ represents a point defect of U(1) spin ice. 
\bea
Q_\tbf{r}=\nabla\cdot S_\tbf{r}
\equiv
\epsilon_\tbf{r}\sum_{\substack{\hat{n}= \hat{x},\hat{y},\hat{z} \\ \eta=\pm 1}}
S_{\tbf{r}+ \eta \hat{n}/2},
\label{eq:Q_def}
\eea
where the coarse-grained spin variable $S_{i=\tbf{r}+\eta\hat{n}/2}$ resides on the cubic link labelled as $i$.
The sign $\epsilon_\tbf{r}\equiv(-1)^{r_x+r_y+r_z}$ defines the local axis of $S_i$ emanating from even ($\epsilon_\tbf{r}=+1$) to odd ($\epsilon_\tbf{r}=-1$) sites. 
With the Gauss law in electromagnetism, the Helmholtz decomposition can be applied, $S_i=S_i^{(L)}+S_i^{(T)}$. 
Only longitudinal modes $S_i^{(L)}$ contribute finite $Q_\tbf{r} =\nabla \cdot S_{\tbf{r}}^{(L)}$ while the transverse modes $S_i^{(T)}$ constitute closed loops of same polarizations of $S_i$, i.e. $\nabla \cdot S_\tbf{r}^{(T)}=0$.
In Supplementary Materials \cite{Supp_domain}, the decomposition is applied to derive the Coulomb interaction Eq. (\ref{eq:Q_eff}) in the longitudinal sector.

For $\mathcal{R}\gg \mathcal{K},\mathcal{U}$, the longitudinal modes are energetically suppressed leading to the spin ice physics, $Q_\tbf{r}=0$ for all $\tbf{r}$. 
This local constraint is relaxed in the presence of defect interactions beyond the on-site.
A locally preferred structure of defect population is developed, which involves both types of modes in $S_i$.
Importantly, the spatial structure and dynamical processes are strongly constrained by the \textit{global neutrality},
\bea
\sum_{\tbf{r}}Q_\tbf{r}=0,
\label{eq:Q_total}
\eea
so that the diffusion-annihilation process of defects always comes with a pair of opposite charges (Fig. \ref{fig:Q_pair}).
This is a consequence of fractionalization, Eq. (\ref{eq:Q_def}), and holds regardless of control parameters and temperature. 

\begin{figure*}[t!]
\subfloat[]
{\label{fig:diagram}
\includegraphics[width=0.34\textwidth]{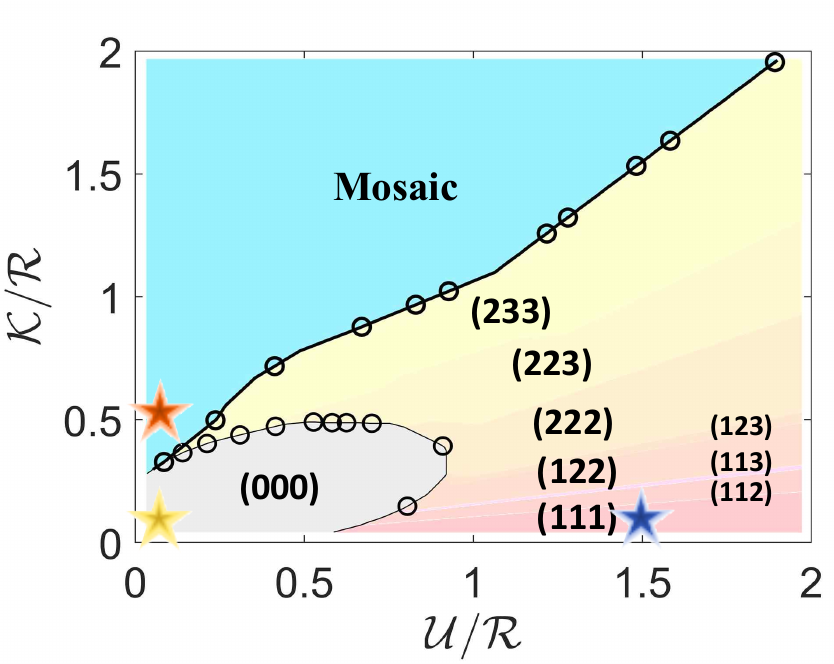}}
\subfloat[]
{\label{fig:ansatz}
\includegraphics[width=0.28\textwidth]{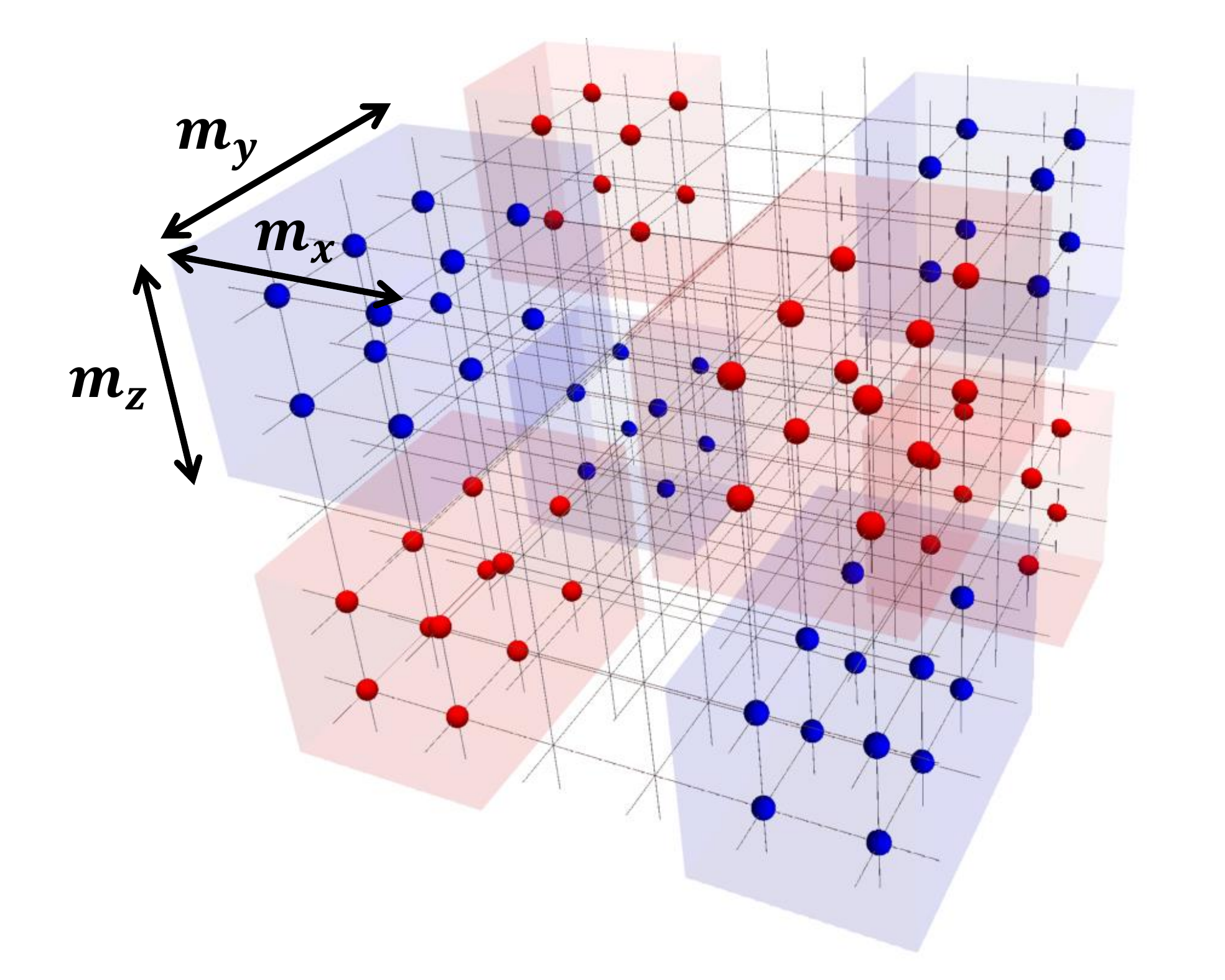}}
\subfloat[]
{\label{fig:auto}
\includegraphics[width=0.35\textwidth]{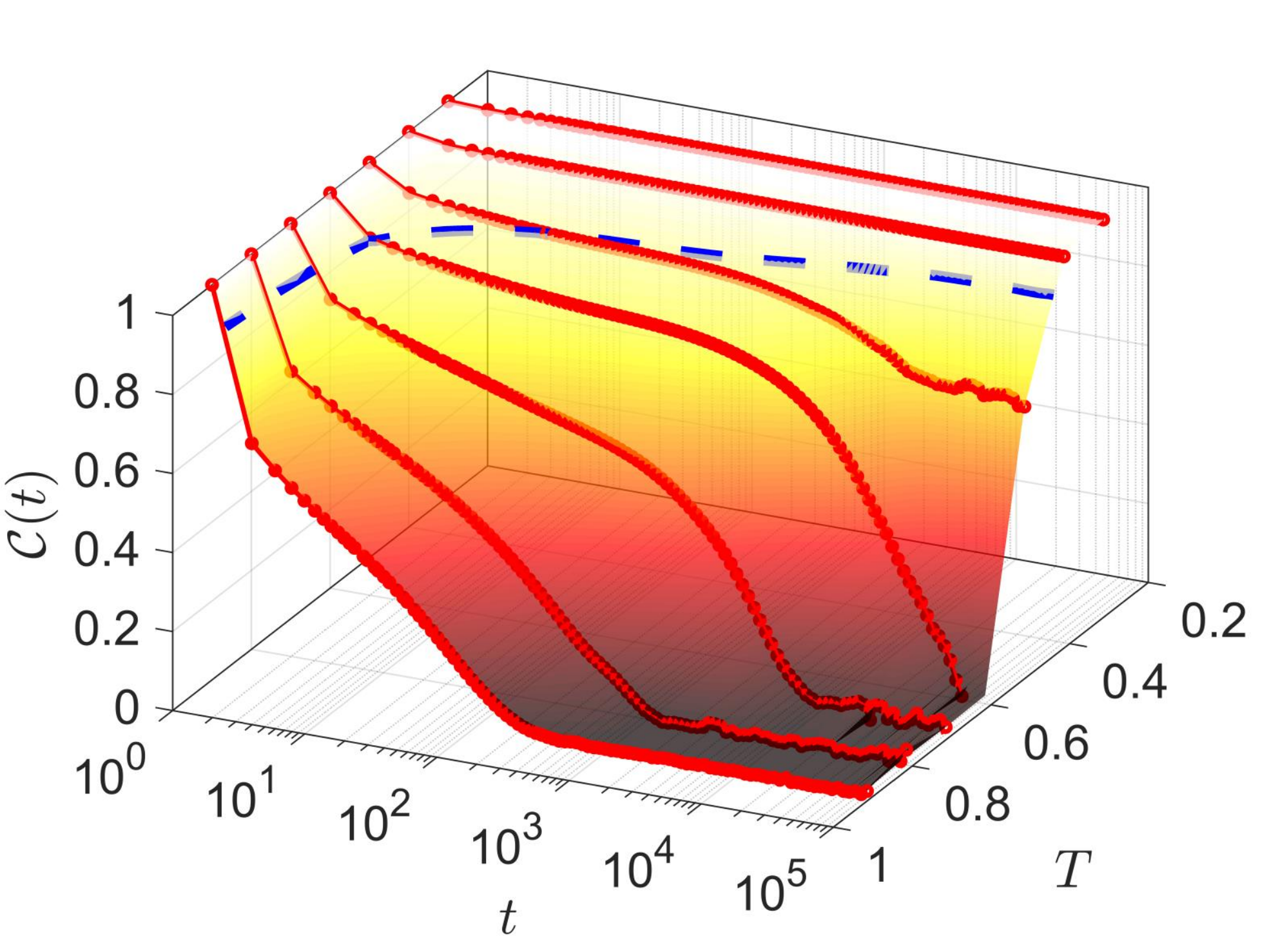}}
\caption{
(a) Ground state phase diagram of Eq. (\ref{eq:Q_eff}) within $0\leq m_{\alpha} \leq m_{\text{max}}=3$.  
Different size of clusters $(m_x,m_y,m_z)$ are stabilized in 9 regimes below the critical line (thick line), $\mathcal{K}<\mathcal{K}_c$. 
while the size infinitely increases above $\mathcal{K}>\mathcal{K}_c$ with the upper bound $m_{\text{max}}$.
The distinct features in charge correlation behaviour are described in (i) Dilute gas (yellow star), (ii) Periodic charge order (blue star), (iii) Mosaic structure (brown  star).
(b) Graphical representation of block-type charge clusters of size $(m_x\times m_y \times m_z)$.
The finite charges $Q_{\tbf{r}}=+1 ~(-1)$ are marked as red (blue) spheres at cubic sites.
A positively charged cluster is surrounded by negatively charged clusters of same sizes, and vice versa. 
(c) Auto-correlation function of charges, $\mathcal{C}_Q(t)$ with normalization $\mathcal{C}_{Q}(t=0)=1$ evaluated at $\mathcal{K}=0.5, \mathcal{U}=0.01$ as the temperature increases from $T=0.3$ to $0.9$ in unit of $\mathcal{R}=1$. 
The system size is $N_{\text{site}}=10^3$ and each sweep contains $N_{\text{site}}$-times of pair-creation/annihilation processes.
The temperature dependence of relaxation time $t_{\text{relax}}$ where $\mathcal{C}_Q(t_{\text{relax}})/\mathcal{C}_Q(t=0)=0.9$ is plotted as the blue dotted line.
}
\label{fig:modulation}
\end{figure*}

\label{sec:Ground state phase diagram}
{\textbf{\textit {Ground state phase diagram ---}}}
Depending on the interaction strengths, the ground states of Eq. (\ref{eq:Q_eff}) exhibit various morphologies of microphase separation under the global neutrality condition, Eq. (\ref{eq:Q_total}). 
When $\mathcal{R}  = 0$, the pair-creation process occurs without on-site cost and most sites are occupied by charged defects, $Q_{\tbf{r}} \neq 0$. 
As the short-range attraction $\mathcal{K}/\mathcal{U}$ increases, the cluster size grows by assembling defects \cite{PhysRevLett.72.1918}.

When $\mathcal{R}> 0$, the domain patterns are more complex and generically organized by both charged and empty sites.
In Fig. \ref{fig:diagram}, the ground state phase diagram exhibits the domain patterns consist of three species, $Q_{\tbf{r}}=-1, 0, +1$ for interaction parameters $\mathcal{K}/\mathcal{R}$ and $\mathcal{U}/\mathcal{R}$.
In each region, the equilibrium pattern is obtained by minimizing Eq. (\ref{eq:Q_eff}) with respect to the cluster size $(m_x, m_y, m_z)$ (Fig. \ref{fig:ansatz}). See the Supplementary Materials for computational details \cite{Supp_domain}.

Likewise the case $\mathcal{R}=0$, the cluster size, $m_\alpha$ increases as $\mathcal{K}/\mathcal{U}$ increases.
For clarity, we set the upper bound of the cluster size to be $m_{\text{max}}=3$ and find 10 different domain patterns within $0\leq m_{\alpha =x,y,z} \leq m_{\text{max}}$. 
The large-$\mathcal{K}$ patterns beyond the range are lumped together with the largest cluster region, $m_\alpha = m_{\text{max}}$ above the critical line.
Above the critical line, the correlation length along with the cluster size constantly increases as the upper bound rises, which implies the emergence of intermediate length scale at some $m_\alpha = m_{\text{max}}$.

The spatial structure of complex domain patterns can be investigated by magnetic correlations. 
We first discuss the defect charge correlations of domain patterns exhibited in Fig. \ref{fig:diagram} by sorting them into
(i) Dilute gas of defects ($m_x=m_y=m_z=0$), 
(ii) Periodic charge orders ($0<m_{\alpha}\leq m_{\text{max}}$) but not all $m_{\alpha}=m_{\text{max}}$, and
(iii) Mosaic structure ($m_x=m_y=m_z=m_\text{max}$) of closely-packed clusters. 
Then static spin structures are also calculated, which straightforwardly identify the charge correlations and detectable in elastic neutron scattering experiments.

In the weakly interacting limit $\mathcal{R}\gg \mathcal{K},\mathcal{U}$, the charge cost penalizes the macroscopic occupation of defects, which reproduces the spin ice, $m_x=m_y=m_z=0$. 
Similar to a dilute gas, the charge correlation is almost negligible, 
\bea
\la Q_\tbf{r} Q_{\tbf{r}'}\ra \approx \delta_{\tbf{r}\tbf{r}'}-\frac{1}{N_{\text{site}}},
\label{eq:Q_corr1}
\eea
where $N_{\text{site}}$ is the total number of cubic sites.
The global neutrality, Eq. (\ref{eq:Q_total}) is ensured by the second term, which is the probability of finding an opposite charge, $Q_{\tbf{r}'}=-1$ at arbitrary sites given $Q_{\tbf{r}}=+1$. 
The weakness of charge correlation is resulted from the energetic suppression of longitudinal modes in spin correlations, which is clarified later. 
Nonetheless, the spin correlation is still non-trivial exhibiting the pinch point singularities in momentum space \cite{PhysRevLett.93.167204,PhysRevB.71.014424, PhysRevLett.110.107202,Henley2010}.

When the interactions $\mathcal{U}/\mathcal{R}$ and $\mathcal{K}/\mathcal{R}$ become significant, a macroscopic amount of defects is populated to develop locally preferred clusters, $m_{\alpha} \neq 0$.
If the long-range frustration is most dominant $\mathcal{U} \gg \mathcal{K},\mathcal{R}$, the nearest-neighbour repulsion strength $\mathcal{U}$ between same-charge defects is much larger than $\mathcal{K}$.
The positive and negative charges are aligned one by one, i.e.
\bea
\la Q_{\tbf{r}}Q_{\tbf{r}'}\ra \approx \epsilon_{\tbf{r}}\epsilon_{\tbf{r}'},
\label{eq:Q_corr2}
\eea
with $m_x=m_y=m_z=1$, and the charge correlation is sharply peaked at $\textbf{R}=\pi(1,1,1)$ in momentum space.
As the short-range attraction $\mathcal{K}$ increases, the correlated cluster gradually grows and charge correlation peak is shifted from $\textbf{R}$. 
In Fig. \ref{fig:diagram}, a succession of lamellar phases, $m_x=m_y=1, m_z>1$ appears in narrow regions above the region of $m_x=m_y=m_z=1$. 
When lamellar length reaches $m_z=m_{\text{max}}$, then $m_x$, $m_y$ grow to develop block-type clusters. 

Above the critical value, $\mathcal{K}>\mathcal{K}_c(\mathcal{U})$, the block-type clusters are densely packed forming a mosaic of domains.
Although the typical sizes of clusters are much smaller than the macroscopic scale, the mosaic structure is more rigid than the vapour-like phase of $\mathcal{R} \gg \mathcal{K}, \mathcal{U}$. 
Despite the absence of long-range order, each single charge constituting the cluster is robustly frozen at low temperature.
To verify the existence of this intermediate order, we investigate the relaxation dynamics of individual charges.

\label{sec:Relaxation dynamics}
{\textbf{\textit {Relaxation dynamics---}}}
To examine the relaxation behaviour, we perform the standard Monte-Carlo (MC) simulation of Eq. (\ref{eq:Q_eff}). 
Due to the global neutrality, the Metropolis dynamics involves the pair-creation or annihilation process at arbitrary two sites $\tbf{r}$ and $\tbf{r}'$ (Fig. \ref{fig:Q_pair}).
In spin model, this process is equivalent to manipulating the Dirac string of $S_i^{(L)}$ connecting $\tbf{r}$ and $\tbf{r}'$ (Fig. \ref{fig:Q_pair}).
In Fig. \ref{fig:auto}, the auto-correlation function $\mathcal{C}_Q(t)$ of defect charges is evaluated in the mosaic structure,
\bea
\mathcal{C}_Q(t)=\frac{1}{N_\text{site}}\la\la\sum_{\tbf{r}}
Q_\tbf{r}(t_0+t)Q_\tbf{r}(t_0)\ra\ra,
\label{eq:auto_corr}
\eea
where $Q_{\tbf{r}}(t=0)$ is the charge distribution of initially equilibrated state at finite temperature $T$ and $\la\la ...\ra\ra$ is the thermal expectation averaged over the MC time, $0\leq t_0 \leq t_{\text{max}}-t$ with $t_{\text{max}}=10^6$ sweeps.

The dynamical quantity, Eq. (\ref{eq:auto_corr}), estimates the correlation of a single charge $Q_\tbf{r}$ at two different times on average.
In the absence of clustering, the initial charge distribution is short-lived and $\mathcal{C}_{Q}(t)$ exponentially decays.
This behaviour is observed in weakly interacting region, such as disordered phase ($T\gg \mathcal{R},\mathcal{K},\mathcal{U}$) and dilute gas ($\mathcal{R}\gg \mathcal{K},\mathcal{U}, T$).
In contrast, for large $\mathcal{K}$, a single charge inside the correlated cluster persists its initial value over a long time and its relaxation can not be fitted by a simple exponential.
The relaxation involves a reconstruction of mosaic domains, in which the interior of cluster is less deformable than the interface due to the energy barrier. 
As the temperature is lowered below $T \sim \mathcal{K}$, the density of deformable interface is rapidly reduced, roughly as the inverse of typical cluster sizes.
Moreover, the global neutrality also restricts the flippable chance per MC step since the flip necessarily takes place at two sites $\tbf{r}$ and $\tbf{r}'$ simultaneously.
In other words, along with the growth of individual clusters, the relaxation is intensely slowed down due to the interplay of energy barrier and global neutrality.

In MC simulation, we find that the plateau is stretched out in intermediate times after a short-time decaying. 
For $\mathcal{K}/\mathcal{R}=0.5$, the plateau becomes apparent below $T/\mathcal{R} \sim 0.7$ and the relaxation proceeds in two-step decaying. 
When the temperature is lowered below $T/\mathcal{R} \sim 0.5$, a major portion of defect charge keeps the initial distribution and the plateau at $\mathcal{C}_Q(t)\gtrsim 0.99$ is maintained over a long time.
The relaxation time, $t_{\text{relax}}$, characterizing the short-time decaying $\mathcal{C}_{Q}(t_{\text{relax}})=0.9$ exhibits a rapid increase below $T/\mathcal{R} \sim 0.5$, which signifies the full covering of locally correlated clusters.

\label{sec:Static magnetic correlation }
{\textbf{\textit {Static magnetic correlation ---}}}
The correlations of domain patterns in Fig. \ref{fig:diagram} can be detected in the spin structure.
We construct the microscopic spin Hamiltonian whose point-like defects interact as Eq. (\ref{eq:Q_eff}), then investigate their static correlations based on large-$N$ approach \cite{PhysRevLett.95.217201,PhysRevLett.110.107202}.
Using Eq. (\ref{eq:Q_def}), the spin Hamiltonian, $H_{\text{spin}}$ can be written as a polynomial of 
\bea
&&
V_{0}=
\begin{pmatrix}
c_x & 2cc_{xy} & 2cc_{xz} \\
2cc_{xy} & c_y & 2cc_{yz} \\
2cc_{xz} & 2cc_{yz} & c_z
\end{pmatrix}
,\;
\label{eq:V_0}
\eea
on the basis of $S_{n=x,y,z}(\tbf{k})$ and 
$c_{\alpha}\!=\!\cos(k_\alpha)$, $cc_{\alpha\beta}\!=\!\cos(\frac{k_\alpha}{2})\cos(\frac{k_\beta}{2})$
in $\tbf{k}$-space.  
Then,
\bea
H_{\text{spin}}=\sum_\tbf{k}
(\mathcal{R}V_{\text{on-site}}+\mathcal{K}V_{\text{attr}})_{\alpha\beta}S_\alpha(-\tbf{k})S_\beta(\tbf{k}),
\label{eq:spin_Hamk}
\eea
where $V_{\text{on-site}}=V_{0}+I_{3\times 3}$, $V_{\text{attr}}=2V_{0}^2-3V_{0}-4I_{3\times 3}$ coincides the first and second terms in Eq. (\ref{eq:Q_eff}) and $I_{3\times 3}$ is the identity matrix.
In Supplementary Materials \cite{Supp_domain}, the explicit form of  $V_0, V_{\text{on-site}}, V_{\text{attr}}$ in Eqs. (\ref{eq:V_0}) and (\ref{eq:spin_Hamk}) are derived on the cubic lattice.
Here, the spin length constraint is imposed by adding Lagrange multiplier, $\mathcal{U}\delta_{\alpha\beta}S_{\alpha}(-\tbf{k})S_{\beta}(\tbf{k})$.
By integrating out the longitudinal modes, $S_i^{(L)}$, the spin stiffness $\mathcal{U}>0$ results in the long-range Coulomb interaction in Eq. (\ref{eq:Q_eff}).

\begin{figure}[t!]
{\label{fig:Spin1}
\includegraphics[width=0.49\textwidth]{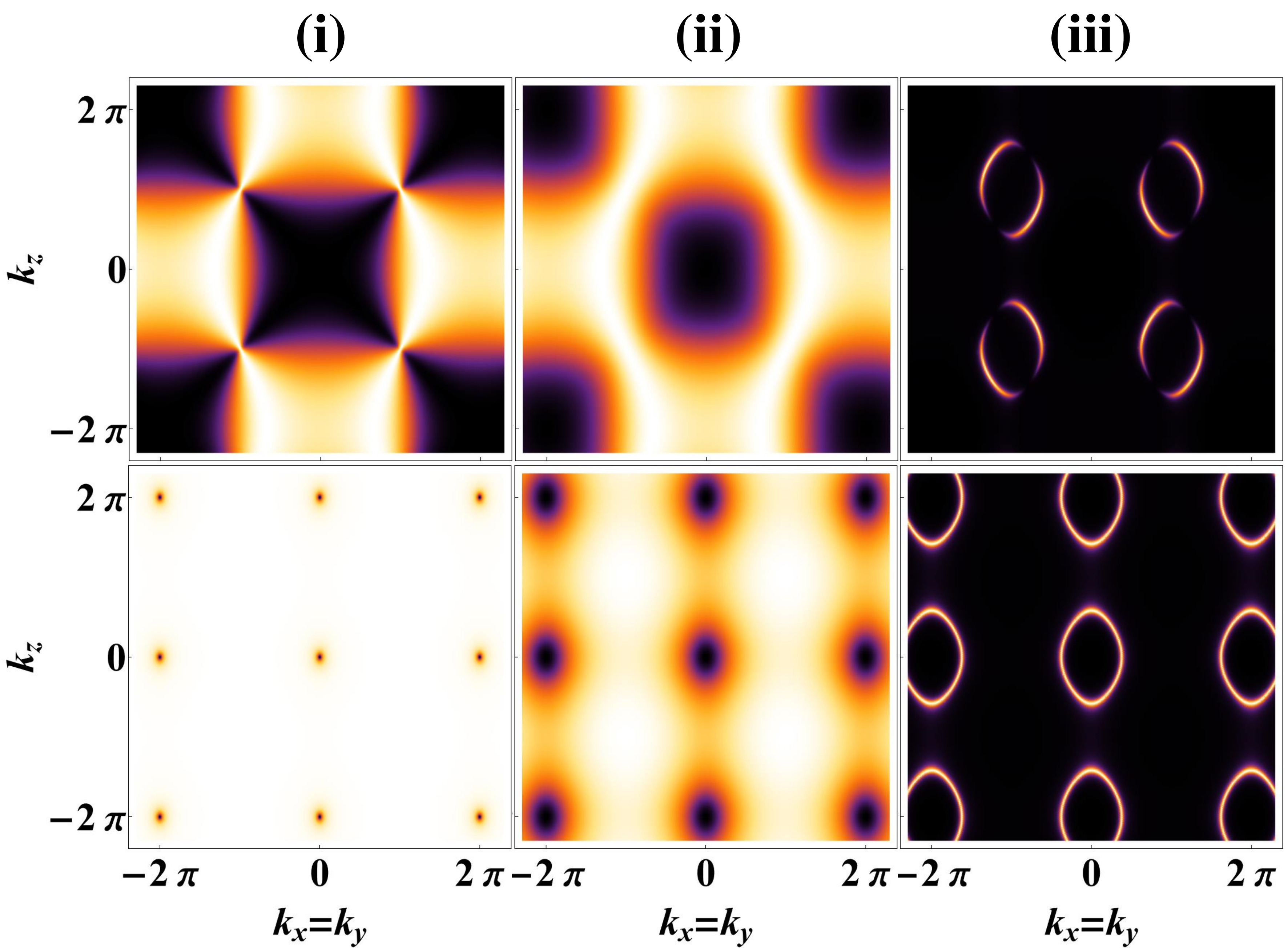}}
\caption{
Static spin correlations $\mathcal{S}_s(\tbf{k})/T$ (upper) and charge correlations $\mathcal{S}_Q(\tbf{k})/T$ (lower) calculated on the $[hhl]$--plane in 
(i) Dilute gas phase, 
(ii) Periodic charge order of defect clusters, 
and 
(iii) Mosaic structure, from left to right.
The intensity is indicated by brightness in arbitrary units. 
The charge correlations exhibit distinct features, (i) featureless, (ii) Charge ordering peak at $\tbf{R}$, (iii) Isotropic ring shape.}
\label{fig:correlation}
\end{figure}

In Fig. \ref{fig:correlation}, the static spin and charge correlations are calculated on the $[hhl]$-plane where
\bea
&&
\mathcal{C}_{S}(\tbf{k})=\sum_{\alpha,\beta = x,y,z}\la S_\alpha(\tbf{k})S_\beta(-\tbf{k})\ra,
\label{eq:spin_corr}
\eea
and
\bea
&&
\mathcal{C}_{Q}(\tbf{k})=\la Q(\tbf{k})Q(-\tbf{k})\rangle.
\label{eq:static_corr}
\eea
In the (i) dilute gas phase, the spin correlation exhibits the sharp pinch point at $\tbf{R}$,
\bea
\la S_{\alpha}(\tbf{R}+\tbf{q})S_{\beta}(\tbf{R}-\tbf{q})\ra \approx \Big(\delta_{\alpha\beta}-\frac{(\tbf{q}\cdot\hat{\alpha})(\tbf{q}\cdot\hat{\beta})}{|\tbf{q}|^2} \Big),\;
\label{eq:spin_corr1}
\eea
as expected in U(1) spin ice.
The spin correlation at $\tbf{R}$ is closely related to the long-range behaviour of defect populations, i.e. the charge correlation at $\boldsymbol{\Gamma}=(0,0,0)$.
Using Eq. (\ref{eq:spin_corr1}), the behaviour Eq. (\ref{eq:Q_corr1}) can be confirmed.
\bea
\mathcal{C}_Q(\tbf{k})\approx 1-\delta_{\tbf{k}\boldsymbol{\Gamma}},
\label{eq:charge_corr1}
\eea
which signifies neither long-range charge order nor defect clustering.
We emphasize that $\mathcal{C}_Q(\boldsymbol{\Gamma})=0$ is always guaranteed by the global neutrality.

In (ii) periodic ordered phases, the pinch point singularity at $\mathcal{C}_S(\tbf{k}= \tbf{R})$ is smoothed to be analytic with a broaden width, which signifies the substantial weight of longitudinal modes in magnetic correlations.
Instead, $\mathcal{C}_Q(\tbf{k})$ is peaked close to $\tbf{R}$ for large $\mathcal{U}$ which characterizes the periodic order of the defect population.

In (iii) mosaic structure, the spin correlation features a half-moon shape close to $\tbf{R}$ as a result of anomalous kinetic term in Eq. (\ref{eq:spin_Hamk}).
The physical meaning of half-moon becomes clear in the charge correlation.
As $\mathcal{K}/\mathcal{U}$ increases from large $\mathcal{U}$ region, the peak in $\mathcal{C}_Q(\tbf{k})$ moves from $\tbf{R}$ and eventually features a sharp ring around $\boldsymbol{\Gamma}$. 
From Eq. (\ref{eq:Q_eff}), the characteristic momentum scales as $|\tbf{k}_c| \sim (\mathcal{U}/\mathcal{K})^{1/4}$ for large $\mathcal{K}$.
As long as the long-range frustration is finite, $\tbf{k}_c$ never reaches $\boldsymbol{\Gamma}$, which agrees with the prohibition of uniform order. 
Here, the charge correlation only settles the characteristic scale in isotropic direction contrary to the point-peak in the large-$\mathcal{U}$ charge ordering phases.


\label{sec:Discussion}
{\textbf{\textit {Discussion ---}}}
We investigate the fractionalized charge clustering as a consequence of competing spin interactions in U(1) quantum spin liquids. 
The long-range defect interaction is inevitable in fractionalized phases and is crucial to both the static and dynamical properties of defect population.
The spatial structure of defect population is constrained by the global neutrality, which leads to novel magnetic correlations.
While the charge correlation is featureless in spin ice, the domain patterns manifest the characteristic features in correlations.
Among them, we confirm the dynamical slowing down in the mosaic structure, which implies a collective behaviour even in the absence of long-range order.
By comparing the static correlations, it turns out that the half moon shape in spin structure manifests the liquid-like order with the intermediate length scale.


Our study signifies that the fractionalized defects in spin liquids exhibit the spatial inhomogeneity and complex domain patterns without disorder.
The block-type clusters in our estimate evolve from the charge ordering to the liquid-like correlation, where the latter region tentatively agrees with the analysis in kagome \cite{PhysRevLett.119.077207} and pyrochlore lattices \cite{Rau2016, PhysRevB.98.144446}.
Both the emergence of intermediate length scale and translational symmetry breaking characterize rich sets of classical spin liquids in addition to spin ice.
Then, the presence of quantum fluctuation would settle novel types of linear superposition in the wave-function beyond the typical ring exchange. 
At the same time, the coexistence of defect population and propagation is also expected, resulting in new types of quantum spin liquids \cite{PhysRevB.104.L100403}.
In addition to novel spatial structures, the defect population offers opportunities to explore the glassy behaviour \cite{PhysRevE.65.065103,PhysRevB.64.174203,PhysRevLett.85.836, PhysRevE.69.021501} such as the tunable fragility and super-Arrhenius relaxation born out of the Coulomb phase. 
We expect that the dynamical properties of domain patterns might be an interesting future work to connect the seemingly independent phenomena of spin liquids and glass physics.

\label{sec: Acknowledgments}
{\textbf{\textit {Acknowledgments ---}}}
H.-J.Y. and S.B.L. acknowledge the support from National Research Foundation (NRF) Grant No. 2021R1A2C1093060. E.-G.M. acknowledges the support from the NRF funded by the Ministry of Science and ICT (No. 2021R1A2C4001847, No. 2022M3H4A1A04074153), and National Measurement Standard Services and Technical Services for SME funded by Korea Research Institute of Standards and Science (KRISS -2022 - GP2022-0014).
\bibliography{QSI_slow}

\setcounter{equation}{0}
\setcounter{figure}{0}
\setcounter{table}{0}
\renewcommand{\theequation}{S\arabic{equation}}
\renewcommand{\thefigure}{S\arabic{figure}}

\pagebreak
\newpage

\thispagestyle{empty}
\mbox{}
\pagebreak
\newpage
\onecolumngrid
\begin{center}
  \textbf{\large 
Structural domain patterns in U(1) quantum spin liquids
  \\Supplementary Materials}\\[.2cm]
  
  Hyeok-Jun Yang$^{1}$, Eun-Gook Moon$^{1}$, and SungBin Lee$^1$\\[.1cm]
  {\itshape ${}^1$Department of Physics, Korea Advanced Institute of Science and Technology, Daejeon, 34141, Korea\\
}
(Dated: \today)\\[1cm]
\end{center}
\onecolumngrid
These Supplementary Materials contain the details on I. Effective action in longitudinal sector, II. Variational ground state energy, and III. Microscopic spin Hamiltonian.

\label{sec: SM_Eff}
\section{I. Effective action in longitudinal sector}
Here, we derive Eq. (\ref{eq:Q_eff}), especially the long-range Coulomb interaction between U(1)-charges from the fractionalization, Eq. (\ref{eq:Q_def}) in continuum based on (i) path integral and (ii) projector methods. Then the continuum correlations are explicitly derived for arbitrary $\mathcal{R}, \mathcal{K}, \mathcal{U}$.

The first and second terms in Eq. (\ref{eq:spin_Hamk}) correspond to the on-site cost and nearest neighbour charge attraction in Eq. (\ref{eq:Q_eff}) respectively. 
Then, the action of Eq. (\ref{eq:spin_Hamk}) is
\bea
\frac{1}{\beta}A[S_{\tbf{r}+ \hat{n}/2}^{(L)},S_{\tbf{r}+ \hat{n}/2}^{(T)},Q_{\tbf{r}},\chi_\tbf{r}, \psi_{\tbf{r},\hat{l}}^{(L)},\psi_{\tbf{r}}^{(T)}] 
&=&
\mathcal{H}[Q_\tbf{r}]
+
2\pi \mathcal{U}
\sum_{\tbf{r}}
\sum_{\substack{\hat{n}= \hat{x},\hat{y},\hat{z}}}
\Big(
(S_{\tbf{r}+ \hat{n}/2}^{(L)})^2 +(S_{\tbf{r}+ \hat{n}/2}^{(T)})^2
\Big)
\nn\\
&+&
i\sum_{\tbf{r}}\chi_\tbf{r}
(Q_\tbf{r}-\epsilon_\tbf{r}\sum_{\substack{\hat{n}, \eta=\pm 1}}
S_{\tbf{r}+ \eta \hat{n}/2}^{(L)})
\nn\\
&+&
i\sum_{\tbf{r},\hat{l}}\psi_{\tbf{r},\hat{l}}^{(L)}
\sum_{\hat{m},\hat{n}}
\epsilon_{lmn}(S_{\tbf{r}+ \hat{m}/2}^{(L)}-S_{\tbf{r}+\hat{m}+\hat{n}/2}^{(L)})
+i\sum_{\tbf{r}}\psi_\tbf{r}^{(T)}
\sum_{\substack{\hat{n},  \eta=\pm 1}}
S_{\tbf{r}+ \eta \hat{n}/2}^{(T)},
\quad
\eea
where $\mathcal{H}[Q_\tbf{r}]$ is the functional of $Q_\tbf{r}$,
\bea
\mathcal{H}[Q_\tbf{r}]=\frac{\mathcal{R}}{2}\sum_{\tbf{r}}Q_\tbf{r}^2
-\mathcal{K}\sum_{\la \tbf{r}\tbf{r}'\ra}Q_{\tbf{r}}Q_{\tbf{r}'},
\label{SMeq:Action_Q}
\eea
and the continuous $\chi_\tbf{r}$-field ensures the Gauss law, Eq. (\ref{eq:Q_def}) while $\psi_{\tbf{r},\hat{l}}^{(L)}$ and $\psi_{\tbf{r}}^{(T)}$-fields impose the Helmholtz decomposition,
\bea
S_{\tbf{r}+ \hat{n}/2}
&=&
S_{\tbf{r}+ \hat{n}/2}^{(L)}+S_{\tbf{r}+ \hat{n}/2}^{(T)}.
\label{SMeq:Helmholtz}
\eea 
The Lagrange multiplier $\mathcal{U}$ is included to satisfy the condition, $\frac{1}{3N_\text{site}}\sum_{\tbf{r}}
\sum_{\substack{\hat{n}= \hat{x},\hat{y},\hat{z}}}
\la(S_{\tbf{r}+ \hat{n}/2})^2\ra=1-x$ with the magnetic doping $0\leq x<1$. 
In the main text, it is shifted, $2\pi \mathcal{U} \rightarrow 2\pi \mathcal{U}+|\lambda_{\tilde{\mathcal{H}}}|$ where $\lambda_{\tilde{\mathcal{H}}}$ is the minimum eigenvalues of interaction matrix of $\mathcal{H}[Q_{\tbf{r}}]$ so that $\mathcal{U}>0$ can be tuned to be arbitrarily small for any $\mathcal{R}$ and $\mathcal{K}$.

Aside from the Hamiltonian part, $\mathcal{H}[Q_\tbf{r}]\rightarrow \tilde{\mathcal{H}}[Q(\tbf{x})]$, the continuum action is
\bea
\frac{1}{\beta}A[\tbf{E}^{(L)}(\tbf{x}),\tbf{E}^{(T)}(\tbf{x}),Q(\tbf{x}),\chi(\tbf{x}),\boldsymbol{\Psi}^{(L)}(\tbf{x}),\psi^{(T)}(\tbf{x})] 
&=&
\int d^3x 
\Big[
2\pi \mathcal{U}\Big((\tbf{E}^{(L)})^2+(\tbf{E}^{(T)})^2 \Big)+i\chi
\Big(
Q-\nabla\cdot \tbf{E}^{(L)}
\Big)
\nn\\
&&
+i\boldsymbol{\Psi}^{(L)}\cdot (\nabla \times \tbf{E}^{(L)})
+i\psi^{(T)}\nabla\cdot \tbf{E}^{(T)}
\Big],
\eea
where the position $\tbf{x}$ is implicit in the right-hand side. And $\epsilon_\tbf{r} S^{(L)}_{\tbf{r}+ \hat{n}/2} \rightarrow\tbf{E}^{(L)}(\tbf{x})$ and $\epsilon_\tbf{r} S^{(T)}_{\tbf{r}+ \hat{n}/2} \rightarrow \tbf{E}^{(T)}(\tbf{x})$ are vector variables in continuum, thus the equation of motion reads i.e. $\nabla \times \tbf{E}^{(L)}=0, \nabla \cdot \tbf{E}^{(T)}=0$.
Integrating out $\tbf{E}^{(L)}$-field,
\bea
\frac{1}{\beta}A'[\tbf{E}^{(T)}(\tbf{x}),Q(\tbf{x}),\chi(\tbf{x}),\psi^{(T)}(\tbf{x})]
&=&
\int d^3x
\Big[
2\pi \mathcal{U}(\tbf{E}^{(T)})^2+i\psi^{(T)}\nabla\cdot \tbf{E}^{(T)}
+
\frac{1}{8\pi\mathcal{U}}(\nabla\chi)^2+i\chi Q
\Big],
\eea
since $\int d^3 x \nabla\chi\cdot(\nabla \times \boldsymbol{\Psi}^{(L)}) =0$ and $\boldsymbol{\Psi}^{(L)}$-field is decoupled from others.
Then, integrating out $\chi$-field leads to
\bea
\frac{1}{\beta}A''[\tbf{E}^{(T)}(\tbf{x}),Q(\tbf{x}),\psi^{(T)}(\tbf{x})]
&=&
\int d^3x
\Big[2\pi \mathcal{U}(\tbf{E}^{(T)})^2+i\psi^{(T)}\nabla\cdot \tbf{E}^{(T)}
+
2\pi\mathcal{U} \Big(Q(\nabla^{-2})Q\Big)
\Big],
\label{SMeq:Action_cont}
\eea
where $(\nabla^{-2})$ is the inverse of the Laplacian whose Fourier transform in 3-dimension is the long-range interaction, i.e. $\frac{1}{4\pi |\tbf{r}|}=\int \frac{d^3k}{(2\pi)^3}\frac{1}{|\tbf{k}|^2}e^{i\tbf{k}\cdot \tbf{r}}$. 
In Eq. (\ref{SMeq:Action_cont}), the transverse (of $\tbf{E}^{(T)}, \psi^{(T)}$-fields) and longitudinal sectors (of $Q$-field) are separated and the effective action in the longitudinal sector is 
\bea
\frac{1}{\beta}\mathcal{S}[Q(\tbf{x})]= 
\tilde{\mathcal{H}}[Q(\tbf{x})] + \frac{\mathcal{U}}{2V}\int d^3x \int d^3x' \frac{Q(\tbf{x})Q(\tbf{x}')}{|\tbf{x}-\tbf{x}'|},
\eea
as Eq. (\ref{eq:Q_eff}) ($V$ is the system volume).
Here, the emergence of Coulomb interaction is resulted from the fractionalization, Eq. (\ref{eq:Q_def}) and holds regardless of the specific form of Eq. (\ref{SMeq:Action_Q}).

Meanwhile, the effective action of $\tbf{E}^{(L)}$ and $\tbf{E}^{(T)}$ by integrating out $\chi,\boldsymbol{\Psi}^{(L)},\psi^{(T)},Q$-fields in $\tbf{k}$-space is
\bea
\frac{1}{\beta}\mathcal{A}[\tbf{E}^{(L)}_\tbf{k}, \tbf{E}^{(T)}_\tbf{k}]
&=&
\frac{1}{\beta}\mathcal{A}_L[\tbf{E}^{(L)}_\tbf{k}]
+\frac{1}{\beta}\mathcal{A}_T[\tbf{E}^{(T)}_\tbf{k}],
\label{SMeq:Action_ELET}
\\
\frac{1}{\beta}\mathcal{A}_L[\tbf{E}^{(L)}_\tbf{k}]
&=&
\int \frac{d^3k}{(2\pi)^3}
\sum_{\alpha\beta\gamma\delta}
P_{\alpha\gamma}^{(L)}(\tbf{k})
\Big(
2\pi \mathcal{U}\delta_{\gamma\delta}
+\frac{\mathcal{R}-6\mathcal{K}}{2}k_{\gamma}k_{\delta}
+\frac{\mathcal{K}}{2}|\tbf{k}|^2k_{\gamma} k_{\delta}
\Big) 
P_{\delta\beta}^{(L)}(\tbf{k})
E_{\tbf{k},\alpha}^{(L)}E_{-\tbf{k},\beta}^{(L)}
\nn\\
&=&
\int \frac{d^3k}{(2\pi)^3}
\sum_{\alpha\beta}
\Big(
2\pi \mathcal{U}\frac{1}{|\tbf{k}|^2}
+\frac{\mathcal{R}-6\mathcal{K}}{2}
+\frac{\mathcal{K}}{2}|\tbf{k}|^2
\Big)
k_\alpha k_\beta E_{\tbf{k},\alpha}^{(L)}E_{-\tbf{k},\beta}^{(L)},
\label{SMeq:Action_EL}
\\
\frac{1}{\beta}\mathcal{A}_T[\tbf{E}^{(T)}_\tbf{k}]
&=&
\int \frac{d^3k}{(2\pi)^3}\sum_{\alpha\beta\gamma\delta}
P_{\alpha\gamma}^{(T)}(\tbf{k})
\Big(
2\pi \mathcal{U}\delta_{\gamma\delta}
\Big)
P_{\delta\beta}^{(T)}(\tbf{k})
E_{\tbf{k},\alpha}^{(T)}E_{-\tbf{k},\beta}^{(T)}
\nn\\
&=&
\int \frac{d^3k}{(2\pi)^3}\sum_{\alpha\beta}2\pi \mathcal{U}\Big(\delta_{\alpha\beta}-\frac{k_\alpha k_\beta}{|\tbf{k}|^2}\Big)E_{\tbf{k},\alpha}^{(T)}E_{-\tbf{k},\beta}^{(T)}.
\label{SMeq:Action_ET}
\eea
where
\bea
P_{\alpha\beta}^{(L)}(\tbf{k})=\frac{k_\alpha k_\beta}{|\tbf{k}|^2}, 
\qquad
P_{\alpha\beta}^{(T)}(\tbf{k})=\delta_{\alpha\beta}-\frac{k_\alpha k_\beta}{|\tbf{k}|^2},
\label{SMeq:Projector}
\eea
project $\tbf{E}_\tbf{k}$-field into longitudinal/transverse components respectively.
It is noteworthy that the first term in Eq. (\ref{SMeq:Action_EL}) is same as the last term in Eq. (\ref{SMeq:Action_cont}) with $Q_\tbf{k}=\sum_{\alpha}k_\alpha E_{\tbf{k},\alpha}^{(L)}$.
Using Eqs. (\ref{SMeq:Action_ELET})-(\ref{SMeq:Action_ET}) and the identity,
\bea
\sum_\beta
\Big(\delta_{\alpha\beta}
+
A k_\alpha k_\beta
\Big)
\Big(
\delta_{\beta\gamma}-\frac{k_\beta k_\gamma}{|\tbf{k}|^2 + A^{-1}}
\Big)
=\delta_{\alpha\gamma},
\eea
for an arbitrary constant $A$,
the correlation of $\tbf{E}=\tbf{E}^{(L)}+\tbf{E}^{(T)}$ is
\bea
\mathcal{G}_{\tbf{E}}(\tbf{k})_{\alpha\beta}
=
\la 
E_{\tbf{k},\alpha} E_{-\tbf{k},\beta}
\ra
=
\la 
E_{\tbf{k},\alpha}^{(L)} E_{-\tbf{k},\beta}^{(L)}
\ra
+
\la 
E_{\tbf{k},\alpha}^{(T)} E_{-\tbf{k},\beta}^{(T)}
\ra
=\frac{1}{4\pi \mathcal{U}\beta}\Big(\delta_{\alpha\beta}-\frac{k_\alpha k_\beta}{|\tbf{k}|^2+\frac{4\pi \mathcal{U}}{\mathcal{R}-6\mathcal{K}
+\mathcal{K}|\tbf{k}|^2
}} \Big)
\label{SMeq:G_E}
\eea
Since the local axis in simple cubic lattice is defined in $\epsilon_\tbf{r}=e^{i\tbf{R}\cdot \tbf{r}}$ (see Eq. (\ref{eq:Q_def})), the behaviour of the correlation $\mathcal{G}_{\tbf{E}}(\tbf{k})_{\alpha\beta}$ can be detected close to $\tbf{R}$, $\mathcal{C}_\mathcal{S}(\tbf{R}+\tbf{k})$ in Eq. (\ref{eq:spin_corr}). 
For example, the spin ice limit $\mathcal{R}\gg \mathcal{K},\mathcal{U}$ reproduces Eq. (\ref{eq:spin_corr1}), which is directly obtained from Eq. (\ref{eq:V_0}).
\bea
\mathcal{G}_{\tbf{E}}(\tbf{k})_{\alpha\beta} \approx \frac{1}{4\pi \mathcal{U}\beta}\Big(\delta_{\alpha\beta}-\frac{k_\alpha k_\beta}{|\tbf{k}|^2}\Big).
\label{SMeq:G_E1}
\eea
Likewise usual pinch point singularities, Eq. (\ref{SMeq:G_E1}) depends on relative angles between $k_x,k_y,k_z$ only and there is no dependence on the magnitude $|\tbf{k}|$.
Similarly, the charge correlation in continuum is
\bea
\mathcal{G}_{Q}(\tbf{k})
=\la Q_\tbf{k}Q_{-\tbf{k}}\ra
=\sum_{\alpha\beta}
k_\alpha k_\beta
\mathcal{G}_{\tbf{E}}(\tbf{k})_{\alpha\beta}=\frac{1}{2\beta}\Big(\frac{1}{\frac{\mathcal{R}-6\mathcal{K}}{2}+\frac{\mathcal{K}}{2}|\tbf{k}|^2+\frac{2\pi \mathcal{U}}{|\tbf{k}|^2}}\Big),
\label{SMeq:G_Q}
\eea
which is isotropic in $\tbf{k}$ and detect the charge correlation close to $\boldsymbol{\Gamma}$, $\mathcal{C}_Q(\tbf{k})$ in Eq. (\ref{eq:static_corr}). Likewise Eq. (\ref{SMeq:G_E1}), Eq. (\ref{SMeq:G_Q}) reproduces the featureless correlation, Eq. (\ref{eq:charge_corr1}) for $\mathcal{R}\gg \mathcal{K},\mathcal{U}$,
\bea
\mathcal{G}_{Q}(\tbf{k}=\boldsymbol{\Gamma})=0
,\qquad
\mathcal{G}_{Q}(\tbf{k}\neq \boldsymbol{\Gamma})
\approx \frac{1}{\beta \mathcal{R}}.
\eea
Again, $\mathcal{G}_{Q}(\tbf{k}=\boldsymbol{\Gamma})=0$ is a consequence of the global neutrality.
For large attraction $\mathcal{K}$, the anomalous kinetic term in Eq. (\ref{SMeq:Action_EL}) sets the characteristic length scale $\tbf{k}_c$ so that Eq. (\ref{SMeq:G_Q}) is peaked at the uniform ring of radius $|\tbf{k}_c|$ around $\boldsymbol{\Gamma}$. 
At the same time, the field correlation $\sum_{\alpha\beta}\mathcal{G}_{\tbf{E}}(k_x,k_x,k_z)_{\alpha\beta}$ depends on not only the relative angle $k_z/k_x$ but also its magnitude $|\tbf{k}|$, which leads to the half-moon shape around $\tbf{R}$ in Fig. \ref{fig:correlation}.

\label{sec: SM_Vari}
\section{II. Variational ground state energy}
The phase diagram Fig. \ref{fig:diagram} is calculated based on the ground state energy of the block-type clustering, Fig. \ref{fig:ansatz}. 
In the simple cubic lattice, the periodic configuration is defined by 6 parameters, the linear size of individual block, $(m_x, m_y, m_z)$ and the separation between nearest-neighbouring blocks, $(n_x,n_y,n_z)$ along the $x,y,z$-directions. As a result, the periodicity of the configuration is $(l_x,l_y,l_z)=(2m_x+2n_x,2m_y+2n_y,2m_z+2n_z)$ in $x,y,z$-directions respectively. 
To ensure the commensurability, the linear size of system $L_\alpha$ in $\alpha$-direction ($N_{\text{site}}=L_xL_yL_z$) is assumed to be an integer multiple of $l_\alpha$ for all $\alpha=x,y,z$. 
With these 6 parameters, the variational ground state energy,  $\la \frac{1}{\beta}\mathcal{S}[Q_{\tbf{r}}]\ra|_{m_\alpha,n_\alpha}\equiv E_{\text{total}}$ is estimated as a sum of on-site cost, $E_{\text{on-site}}$, short-range attractive energy, $E_{\text{attr}}$, and the Coulomb energy, $E_{\text{Coulomb}}$ where $\la ...\ra|_{m_\alpha,n_\alpha}$ is the expectation value with respect to the $(m_\alpha, n_\alpha)$-configuration \cite{PhysRevE.62.7781} and,
\bea
E_{\text{total}}[m_\alpha, n_\alpha]
&=&
E_{\text{on-site}}[m_\alpha, n_\alpha]
+
E_{\text{attr}}[m_\alpha, n_\alpha]
+
E_{\text{Coulomb}}[m_\alpha, n_\alpha],
\label{SMeq:E_total}
\\
\frac{1}{N_\text{site}} E_{\text{on-site}}[m_\alpha, n_\alpha]
&=&
\frac{\mathcal{R}}{2}\frac{m_xm_ym_z}{(m_x+n_x)(m_y+n_y)(m_z+n_z)},
\label{SMeq:E_on-site}
\\
\frac{1}{N_\text{site}} E_{\text{attr}}[m_\alpha, n_\alpha]
&=&
-\mathcal{K}\frac{3m_xm_ym_z-(1+\delta_{n_x,0})m_ym_z-(1+\delta_{n_y,0})m_xm_z-(1+\delta_{n_z,0})m_xm_y}{(m_x+n_x)(m_y+n_y)(m_z+n_z)},
\label{SMeq:E_attr}
\\
\frac{1}{N_\text{site}} E_{\text{Coulomb}}[m_\alpha, n_\alpha]
&=&
\frac{2\pi \mathcal{U}}{N_{\text{site}}}\sum_{\tbf{k}}{}^{'}
V(\tbf{k})Q(\tbf{k})Q(-\tbf{k}),
\label{SMeq:E_Coulomb}
\eea
where $Q(\tbf{k})$ and $V(\tbf{k})$ are lattice Fourier transforms of the charge $Q_\tbf{r}$ and the Coulomb interaction $\sim 1/|\tbf{r}|$ respectively. 
\bea
Q(\tbf{k})
&=&
\frac{1}{\sqrt{N_\text{site}}}\sum_{\tbf{r}}Q_{\tbf{r}}e^{i\tbf{k}\cdot \tbf{r}},
\label{SMeq:Q_k}
\\
V(\tbf{k})
&=&
\frac{1}{2\sum_{\alpha =x,y,z}\Big(1-\cos(k_\alpha)\Big)}-v_0,
\label{SMeq:V_k}
\eea
and
\bea
v_0
&=&
\int_{-2\pi}^{2\pi}\frac{dk_xdk_ydk_z}{(2\pi)^3} \frac{1}{2\sum_{\alpha =x,y,z}\Big(1-\cos(k_\alpha)\Big)}\approx 0.253.
\eea 
In Eq. (\ref{SMeq:E_Coulomb}), the summation runs over the subsets of $\tbf{k}$-space satisfying the periodicity of $l_\alpha=2(m_\alpha+n_\alpha)$, i.e. $e^{ik_\alpha (m_\alpha+n_\alpha)}=-1$. 
In other words, for some integer $0\leq p_\alpha \leq m_\alpha+n_\alpha -1$,
\bea
k_\alpha =\frac{\pi}{m_\alpha +n_\alpha}(2p_\alpha +1),
\eea
and $\sum_{\tbf{k}}{}^{'}=\sum_{p_\alpha=0}^{m_\alpha+n_\alpha-1}$.
Meanwhile, Eq. (\ref{SMeq:Q_k}) with discrete values, $Q_\tbf{r}=-1,0,+1$ is written as
\bea
Q(\tbf{k})
&=&
e^{i(k_1+k_2+k_3)}\frac{\sqrt{N_\text{site}}}{(m_x+n_x)(m_y+n_y)(m_z+n_z)}
\frac{(1-e^{ik_xm_x})(1-e^{ik_ym_y})(1-e^{ik_zm_z})}{(1-e^{ik_x})(1-e^{ik_y})(1-e^{ik_z})},
\nn\\
|Q(\tbf{k})|
&=&
\sqrt{N_\text{site}}\prod_{\alpha =x,y,z}
\frac{1}{m_{\alpha}+n_\alpha}
\Big|\frac{\sin(\frac{k_\alpha m_\alpha}{2})}{\sin(\frac{k_\alpha}{2})} \Big|.
\label{SMEq:AbsQ_k}
\eea
Thus, Eqs. (\ref{SMeq:E_on-site})-(\ref{SMeq:E_Coulomb}) result in the variational ground state energy, Eq. (\ref{SMeq:E_total}) as a function of $(m_\alpha, n_\alpha)$. 
For example, minimizing the attractive energy, Eq. (\ref{SMeq:E_attr}) requires the cluster size, $m_\alpha$ to be large as possible.
The phase diagram, Fig. \ref{fig:diagram} is obtained by minimizing Eq. (\ref{SMeq:E_total}) with respect to $(m_\alpha, n_\alpha)$ within $0\leq m_\alpha, n_\alpha \leq 3$ and separates the regime of different cluster sizes.

\label{sec: SM_Spin}
\section{III. Microscopic spin Hamiltonian}
The microscopic spin Hamiltonian is defined in Eq. (\ref{eq:spin_Hamk}) so that its effective action of defects becomes Eq. (\ref{eq:Q_eff}) in the longitudinal sector. 
Here, Eqs. (\ref{eq:V_0}) and (\ref{eq:spin_Hamk}) are explained in detail.

We consider the Lieb lattice on 3-dimension, whose lattice sites consists of full sets of cubic sites and mid-points of cubic links. 
Likewise in main text, our notation labels $\tbf{r}$ and $i\equiv \tbf{r}+\frac{1}{2}\hat{n}\; (\hat{n}=\hat{x},\hat{y},\hat{z})$ for cubic sites and mid-points of cubic links respectively.

\begin{figure}[t!]
{\label{SMfig:Lieb1}
\includegraphics[width=0.3\textwidth]{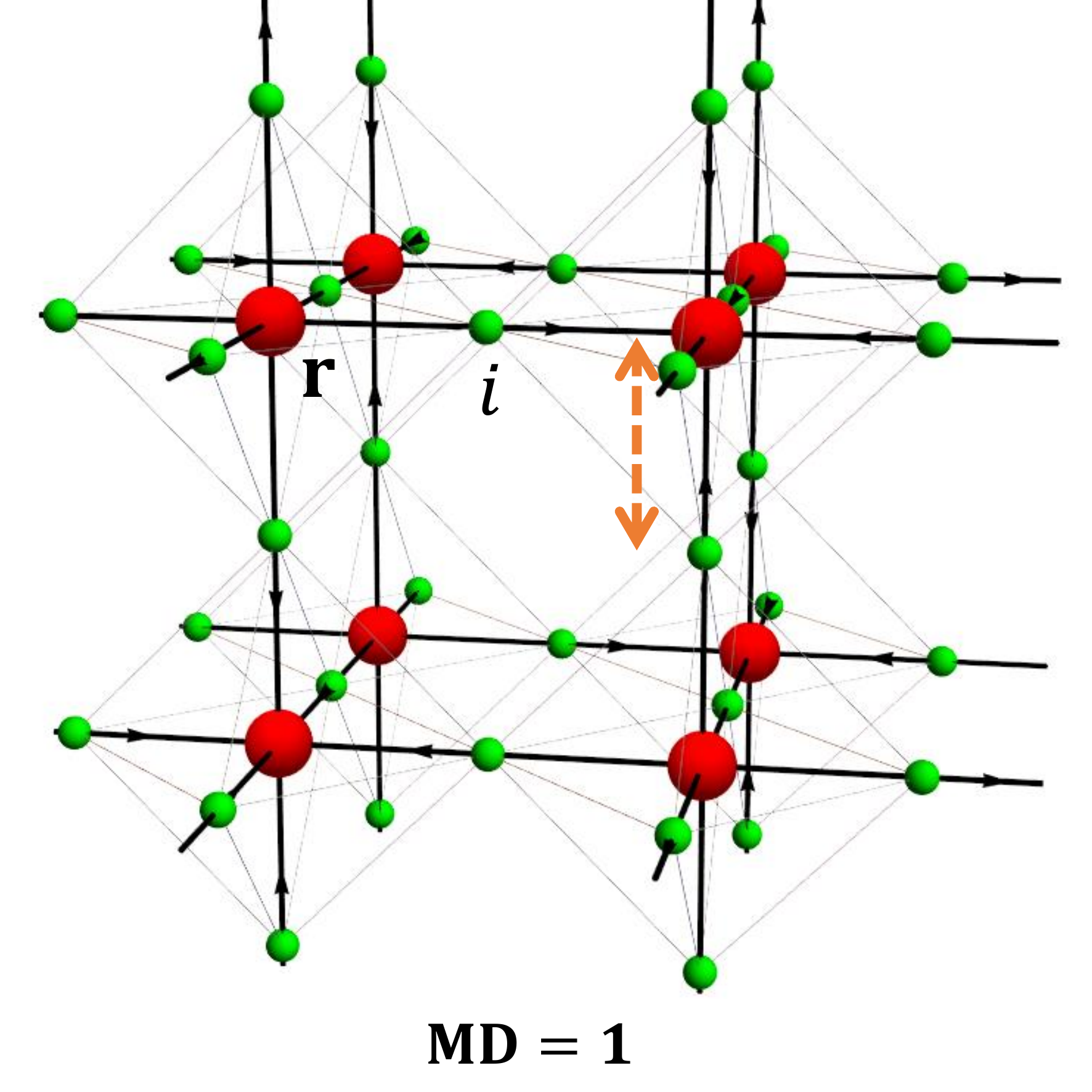}}
\quad
{\label{SMfig:Lieb2}
\includegraphics[width=0.3\textwidth]{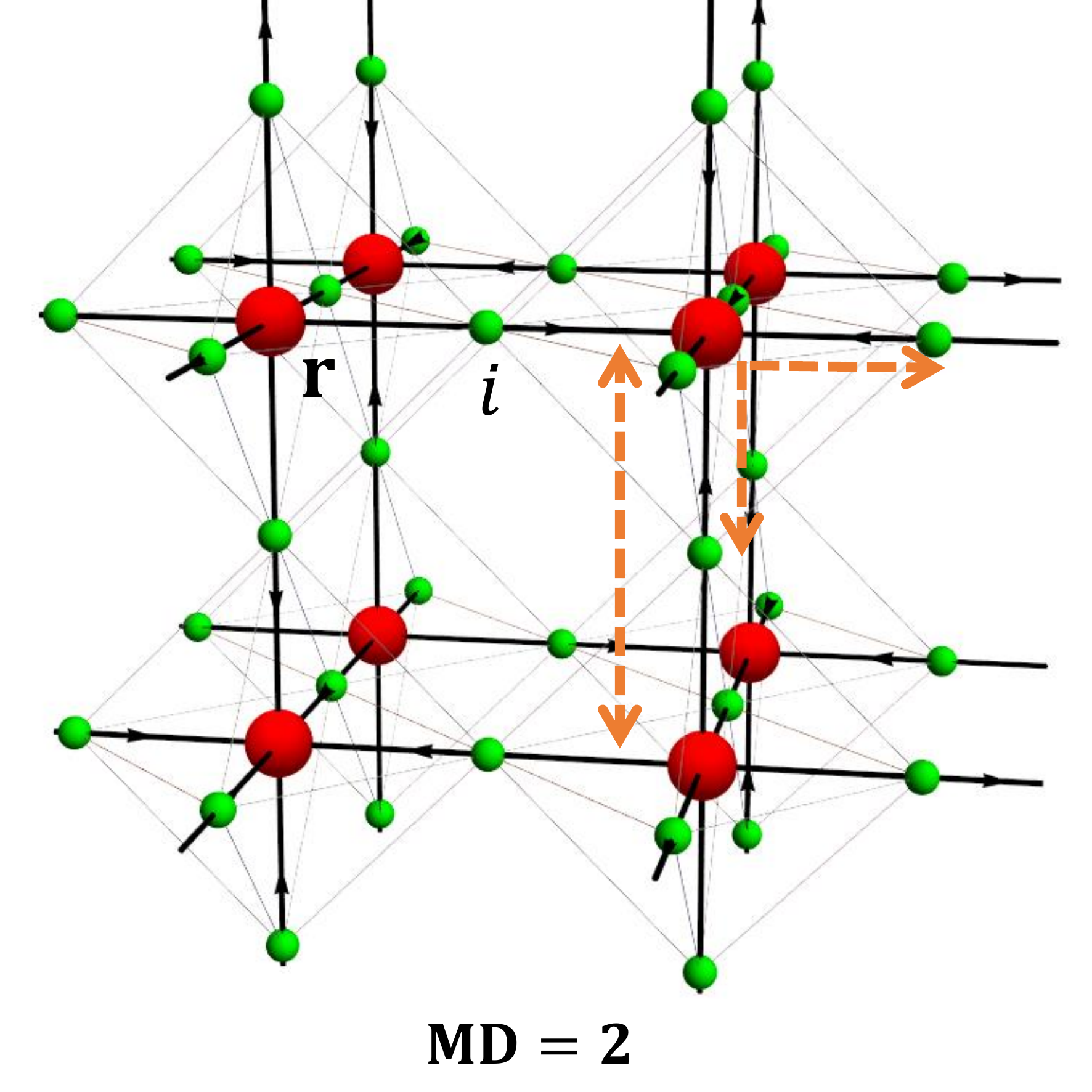}}
\quad
{\label{SMfig:Lieb3}
\includegraphics[width=0.3\textwidth]{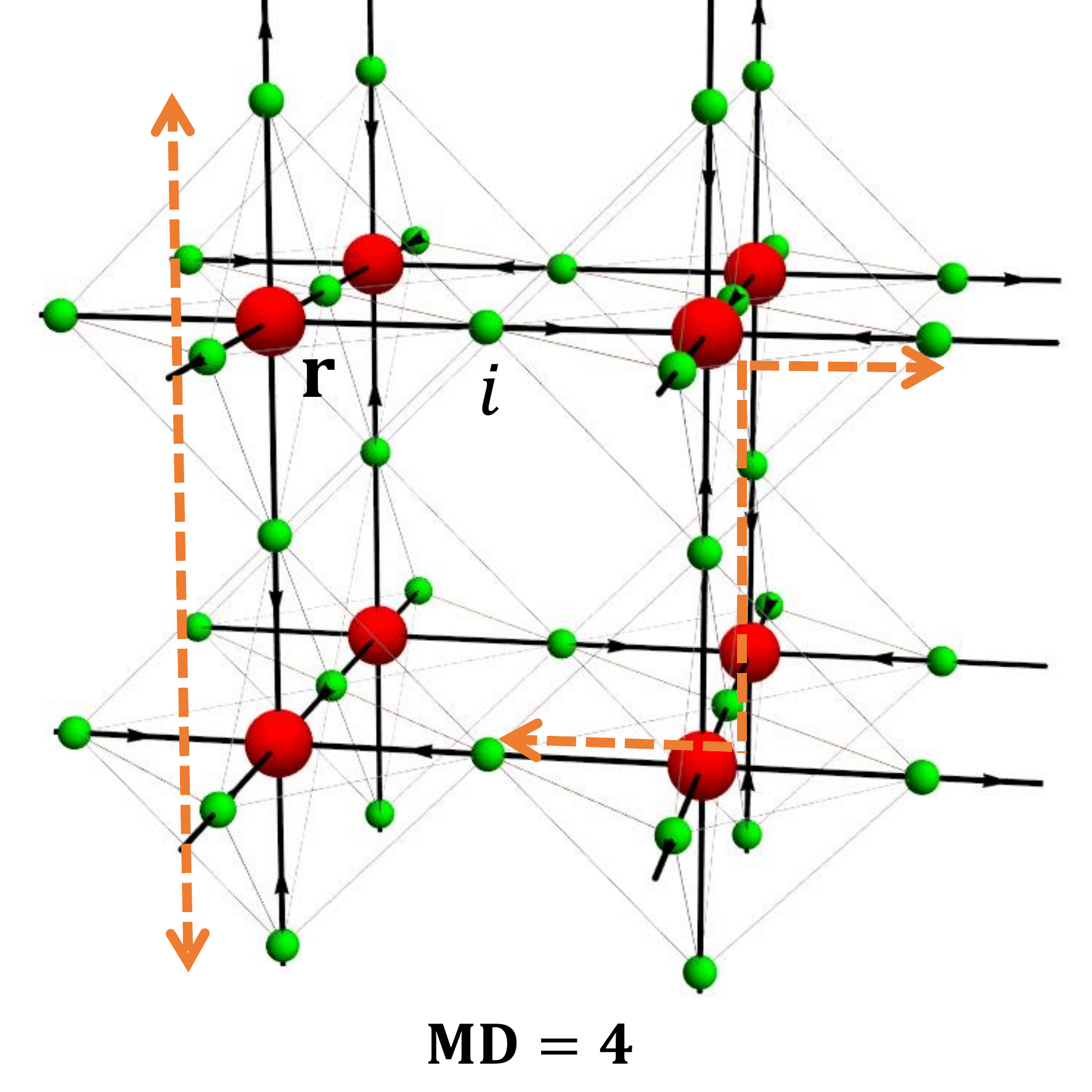}}
\caption{
The 3-dimensional Lieb lattice formed by cubic sites, $\tbf{r}$ (red sphere) and mid-points of links, $i$ (green sphere). 
The mid-point $i$ can be viewed as the vertex of conner-sharing network of octahedron surrounding a cubic site $\tbf{r}$. Similarly, $\tbf{r}$ is an octahedron center. 
Between the center and vertex belong to a octahedron, $\text{MD}=1$. 
Between two vertices belong to a single octahedron, $\text{MD}=2$. 
Between two vertices belong to nearest neighbouring octahedrons, $\text{MD}=4$.
A few examples of paths with $\text{MD}=1$ (left), $\text{MD}=2$ (middle) and $\text{MD}=4$ (right) on the Lieb lattice are shown in dotted arrows.
}
\label{SMfig:Lieb}
\end{figure}

The on-site and nearest neighbour charge interaction in Eq. (\ref{eq:Q_eff}) turn out to be built from the spin interaction matrix whose elements depend on only the Manhattan Distance (MD) \cite{PhysRevB.98.144446} on the Lieb lattice (Fig. \ref{SMfig:Lieb}).
Here, the MD between two Lieb lattice sites is defined as a length of shortest path connecting them along bonds in unit of $\frac{1}{2}\hat{n}$.
For example, the MD between the cubic site $\tbf{r}$ and the mid-point $i$ are always odd, particularly, $\text{MD}=1$ between $\tbf{r}$ and the closest mid-point $\tbf{r}+\frac{1}{2}\hat{n}$, with the interaction matrix in $\tbf{k}$-space, 
\bea
V_{\text{MD}=1}(\tbf{k})
= 2
\begin{pmatrix}
0 & 0 & 0 & \cos(\frac{k_x}{2}) \\
0 & 0 & 0 & \cos(\frac{k_y}{2}) \\
0 & 0 & 0 & \cos(\frac{k_z}{2}) \\
\cos(\frac{k_x}{2}) & \cos(\frac{k_y}{2}) & \cos(\frac{k_z}{2}) & 0 
\end{pmatrix},
\label{SMeq:V_MD1}
\eea
where from the first to third row/columns are assigned for modes on the mid-points $i=\tbf{r}+\frac{1}{2}\hat{n}$ and the fourth is for the cubic site $\tbf{r}$. When the matrix elements of total Hamiltonian depend on MD only, the interaction matrix is expanded as a polynomial of Eq. (\ref{SMeq:V_MD1}).

The MD is always even between mid-points of links or between cubic sites, thus the spin Hamiltonian is represented as a sum of interaction matrices $V_{\text{MD}=n}$ with even $n$.
In other words, the matrix $V_{\text{MD}=n}$ (even $n$) is block-diagonal.
The on-site charge interaction is obtained from the nearest-neighbour spin exchanges, i.e. interaction matrix of $\text{MD}=2$ (Fig. \ref{SMfig:Lieb}), 
\bea
V_{\text{MD}=2}(\tbf{k})
&=&
\Big(V_{\text{MD}=1}(\tbf{k})\Big)^2
-\begin{pmatrix}
2 & 0 & 0 & 0 \\
0 & 2 & 0 & 0 \\
0 & 0 & 2 & 0 \\
0 & 0 & 0 & 6 \\
\end{pmatrix}
\nn\\
&=&
2
\begin{pmatrix}
\cos(k_x) & 2\cos(\frac{k_x}{2})\cos(\frac{k_y}{2}) & 2\cos(\frac{k_x}{2})\cos(\frac{k_z}{2}) & 0 \\
2\cos(\frac{k_x}{2})\cos(\frac{k_y}{2}) & \cos(k_y) & 2\cos(\frac{k_y}{2})\cos(\frac{k_z}{2}) & 0 \\
2\cos(\frac{k_x}{2})\cos(\frac{k_z}{2}) & 2\cos(\frac{k_y}{2})\cos(\frac{k_z}{2}) & \cos(k_z) & 0 \\
0 & 0 & 0 & \cos(k_x)+\cos(k_y)+\cos(k_z)
\end{pmatrix},
\label{SMeq:V_MD2}
\eea
then projecting out the fourth row/column,
\bea
\tilde{V}_{\text{MD}=2}(\tbf{k})
\equiv
P_{i}{V}_{\text{MD}=2}(\tbf{k})P_{i}
=2
\begin{pmatrix}
\cos(k_x) & 2\cos(\frac{k_x}{2})\cos(\frac{k_y}{2}) & 2\cos(\frac{k_x}{2})\cos(\frac{k_z}{2}) \\
2\cos(\frac{k_x}{2})\cos(\frac{k_y}{2}) & \cos(k_y) & 2\cos(\frac{k_y}{2})\cos(\frac{k_z}{2})\\
2\cos(\frac{k_x}{2})\cos(\frac{k_z}{2}) & 2\cos(\frac{k_y}{2})\cos(\frac{k_z}{2}) & \cos(k_z)
\end{pmatrix},
\eea
where $P_i$ is the projector into the subspace of modes on mid-points $i$. 
In the main text, we label $V_0=\tilde{V}_{\text{MD}=2}/2$ in Eq. (\ref{eq:V_0}).
From the definition Eq. (\ref{eq:Q_def}), we have
\bea
&&V_{\text{on-site}}(\tbf{k})=\frac{1}{2}\tilde{V}_{\text{MD}=2}(\tbf{k})+I_{3\times 3}.
\label{SMeq:V_onsite}
\eea
Similarly, the nearest-neighbour charge interaction is obtained by combining $I_{3\times 3}$, $\tilde{V}_{\text{MD}=2}$ and $\tilde{V}_{\text{MD}=4}$, where
\bea
&&
\tilde{V}_{\text{MD}=4}(\tbf{k})
=
\Big(\tilde{V}_{\text{MD}=2}(\tbf{k})\Big)^2
-4\tilde{V}_{\text{MD}=2}(\tbf{k})
-
10I_{3\times 3},
\label{SMeq:V_MD4}
\eea
and
\bea
&&
V_{\text{attr}}(\tbf{k})=\frac{1}{2}\tilde{V}_{\text{MD}=4}(\tbf{k})
+\tilde{V}_{\text{MD}=2}(\tbf{k})
+I_{3\times 3}, 
\label{SMeq:V_attr}
\\
&&
H_{\text{spin}}(\tbf{k})
=\mathcal{R}V_{\text{on-site}}(\tbf{k})
+\mathcal{K}V_{\text{attr}}(\tbf{k}).
\label{SMeq:H_spin}
\eea
The static spin structure $\mathcal{C}_S(\tbf{k})$ in Eq. (\ref{eq:spin_corr}) is obtained by calculating $\Big(2\pi\mathcal{U}I_{3\times 3}+H_{\text{spin}}(\tbf{k})\Big)^{-1}$.
Also,
\bea
Q_{\tbf{r}}
&=&
e^{i\tbf{R}\cdot\tbf{r}}\sum_{\hat{n}=\hat{x},\hat{y},\hat{z}}\Big(S_{\tbf{r}+\frac{\hat{n}}{2}}+S_{(\tbf{r}-\hat{n})+\frac{\hat{n}}{2}} \Big),
\nn\\
Q(\tbf{k})
&=&
\frac{1}{\sqrt{N_{\text{site}}}}\sum_\tbf{r}Q_{\tbf{r}}e^{i\tbf{k}\cdot\tbf{r}}
=
\frac{2}{\sqrt{N_{\text{site}}}}
\sum_{\hat{n}}S_{n}(\tbf{k}+\tbf{R})
\cos\Big((\tbf{k}+\tbf{R})\cdot\frac{\hat{n}}{2}\Big)
=-\frac{2}{\sqrt{N_{\text{site}}}}\sum_{\hat{n}}S_{n}(\tbf{k}+\tbf{R})\sin\Big(\tbf{k}\cdot\frac{\hat{n}}{2}\Big), \quad
\eea
and the charge correlation, Eq. (\ref{eq:static_corr}) is
\bea
\mathcal{C}_Q(\tbf{k})=\frac{4}{N_{\text{site}}}\sum_{\hat{n},\hat{m}}
\la S_{n}(\tbf{k}+\tbf{R})S_{m}(-\tbf{k}-\tbf{R})\ra
\sin(\tbf{k}\cdot\frac{\hat{n}}{2})\sin(\tbf{k}\cdot\frac{\hat{m}}{2}).
\eea
which reduces to Eq. (\ref{SMeq:G_Q}) in the limit, $|\tbf{k}|\rightarrow 0$.

\end{document}